%% file: sgpau.tex
\documentclass[fleqn,usenatbib]{mnras}

\usepackage{newtxtext,newtxmath}
\usepackage{comment}
\usepackage{multicol}
\usepackage{graphicx}
\usepackage{float}
\usepackage{graphicx, subfigure}
\usepackage{listings}
\usepackage{longtable}
\usepackage[toc,page]{appendix}
\usepackage{xspace}

\newcommand{\NOOP}[1]{}

\newcommand{\COMMENT}[1]{\NOOP{#1}}

\newcommand{\NS}[1]{{#1}}
\newcommand{\LC}[1]{{#1}}

\newcommand{\sxt}{{\tt SExtractor}\xspace}
\newcommand{\cnn}{{CNN}\xspace}
\newcommand{\nn}{{ANN}\xspace}
\newcommand{\cnnc}{{\it cnn\_stellarity}\xspace}

\usepackage[T1]{fontenc}
\usepackage{ae,aecompl}

\usepackage{graphicx}	
\usepackage{amsmath}	
\usepackage{amssymb}	

\title[The PAU Survey: star-galaxy classification with multi narrow-band data]{The PAU Survey: star-galaxy classification with multi narrow-band data}

\input{authors.tex}

\date{Accepted XXX. Received YYY; in original form ZZZ}

\pubyear{2018}
\usepackage{hyperref} 
\hypersetup{colorlinks,citecolor=blue,linkcolor=blue,urlcolor=blue}
\hypersetup{draft}
\begin{document}
\label{firstpage}
\pagerange{\pageref{firstpage}--\pageref{lastpage}}
\maketitle
%
\begin{abstract}
Classification of stars and galaxies is a well-known astronomical problem that has been treated using different approaches, most of them relying on morphological information. In this paper, we tackle this issue using the low-resolution spectra from narrow band photometry, provided by the PAUS (Physics of the Accelerating Universe) survey. We find that, with the photometric fluxes from the 40 narrow band filters and without including morphological information, it is possible to separate stars and galaxies to very high precision, $98.4\%$ purity with a completeness of $98.8\%$ for objects brighter than $I=22.5$. This precision is obtained with a Convolutional Neural Network as a classification algorithm, applied to the objects' spectra. We have also applied the method to the ALHAMBRA photometric survey and we provide an updated classification for its Gold sample.
\end{abstract}


\section{INTRODUCTION }
\label{sec:intro}
A basic step in the extraction of astronomical information from photometric images is the separation of stars from galaxies. This is vital in a photometric survey in order to provide pure samples with minimal systematic contribution from the effect of cross-contamination of the star and galaxy samples, to be used for parameter estimation or model comparison in astrophysical or cosmological analyses \LC{(see, e.g., \citet{Soumagnac} to see the impact of this issue in large scale structure and weak lensing and \citet{sgdes}, for the impact in Milky Way studies).}\\ 

Historically, there have been many different approaches to tackle this problem. The first classification methods were morphology based \citep{MacGillivray,Kron,Yasuda,Leauthaud} and they consisted \NS{of} the estimation of an optimal cut on the space of observable image properties, such as a magnitude-size space, or in statistical properties such as measured second-order moments. However, these methods perform poorly when classifying faint objects as morphological information contained in noisy measurements is limited. Improved classification based on morphology relying on more advanced algorithms have been reported in \citet{sgdes,jplussg}.\\

Other classification methods use Bayesian-based approaches \LC{\citep{Sebok, Henrion, Fadely, Rachen}}. The application of a Bayesian classification approach to multi-band data must consider information coming from morphologies and color: the morphological features of a galaxy will be correlated with its magnitude. Hence, as the number of photometric bands increases, this approach gets more and more complicated.\\

Another approach is that provided by machine learning algorithms, which have emerged as an important tool for classification \citep[see e.g. ][]{Odewahn, SExtractor, Soumagnac}. It consists of learning the underlying behavior of a given class sample adaptively from the training data and the later generalization of this learning to samples beyond such training data.\\

Different possible machine learning algorithms are used in star galaxy classification problems, such as boosted decision trees (BDT) \citep{Sevilla-Noarbe}, neural networks (ANN) \citep[e.g.][]{Soumagnac} or random forests (RF) \citep[e.g.][]{Morice-Atkinson}, most of them trained on morphologically-based \COMMENT{truth tables} \LC{flags} from external, deeper datasets or detailed simulations.\\ Most broad band photometric surveys, such as  DES \citep{DES}, SDSS \citep{SDSS} or PANSTARRS \citep{PANSTARRS}, rely on morphological information, with \LC{limited} \COMMENT{just moderate} evidence that this can be improved with color information \citep{sgdes}, without resorting to infrared data \citep{kovacs,banerji}.  \\

For the case of narrow-band data one can ask if it is possible to distinguish stars and galaxies only from the fluxes. This way, narrow-band surveys, which do not go as deep as their broad-band counterparts, could provide an accurate classification based on their flux distribution as well. In this work, we  examine this question considering several machine learning approaches. \\

We will discuss the performance of machine learning algorithms on multiple narrow-band color information using PAUS \citep[Physics of the Accelerated Universe Survey, ][]{pauspie,martipau} and ALHAMBRA \citep[Advanced Large Homogeneous Area Medium Band Redshift Astronomical survey, ][]{ALHAMBRA,Moles}. In the case of PAUS, the classification can be compared with that provided by \sxt \citep{SExtractor}, a software that detects, deblends, measures and classifies sources from astronomical images. \LC{\sxt provides two star-galaxy classifiers: {\tt CLASS\_STAR} and {\tt SPREAD\_MODEL}. The former relies on a multilayer feed-forward neural network trained with 10 inputs: the object's peak pixel value above the local background, 8 isophotal areas and an estimate for the seeing. On the other hand, {\tt SPREAD\_MODEL} indicates whether a model for local PSF or a slightly extended galaxy model best fit the image data.} Concerning ALHAMBRA, we will apply our algorithm and compare with the current classification scheme from the Gold catalog \citep{Molino}, which is based on photometric fluxes and morphologies.  \COMMENT{Lastly, we will also study how the performance of the algorithm scales with the resolution of the spectra, i.e with the number of bands in a given wavelength range, comparing PAUS, ALHAMBRA and the Subaru $g,r,i,z$ bands \citep{taniguchi}.}\\

The standard processing of PAUS images is carried out by performing forced photometry (S.Serrano et al. in prep.): the fluxes from objects are computed at predefined reference positions from external catalogs, in order to obtain more precise photometric redshifts for these broad band detections. \LC{In the case of COSMOS, the external catalog is the COSMOS Photometric Redshift Catalog\footnote{\url{https://irsa.ipac.caltech.edu/data/COSMOS/gator_docs/cosmos_zphot_mag25_colDescriptions.html}} \citep{laigle}}.  Therefore, the objects are already classified from deeper observations, before applying any further method. The results from this paper are meant to demonstrate the efficiency of machine learning algorithms on astronomical classification problems using narrow band photometry spectra, and may be useful to think of implementation of these algorithms to other crucial issues, such as galaxy classification, photo-z or outlier rejection for this kind of data. In addition, the objects used for PAUS photometric calibration are SDSS stars (Castander et al. in prep.), so it would be of great interest for PAUS to have its own classified stars to perform such a calibration, with a more pure selection. Lastly, an algorithm able to classify objects with low-resolution spectra would also be interesting for planned or future narrow-band surveys. \COMMENT{such as future extensions of PAUS, J-PLUS \citep{jplus} and J-PAS \citep{J-PAS}}\\

The layout of this paper is as follows; in section \ref{sec:data} we will define the datasets employed in the analysis. In section \ref{sec:method}, there is a short definition of all the machine learning algorithms used at some point in this study. The characterization of the algorithms is done in section \ref{sec:characterization} and section \ref{sec:results} provides the results on the ALHAMBRA and PAUS datasets. In section \ref{sec:conclusions} there is a final discussion of the main results. \\

\section{Data}
\label{sec:data}
In this work we would like to assess the performance of a  machine learning  classifier over two narrow-band datasets, PAUS early data and the ALHAMBRA Gold catalog\footnote{\url{https://cloud.iaa.csic.es/alhambra/}}, in the latter case comparing with the standard classification provided by that survey. We will work on the COSMOS field\footnote{\url{http://cosmos.astro.caltech.edu/}} comparing against the COSMOS space-based imaging catalog \citep{Leauthaud}, which provides a morphology-based classification (\texttt{MU\_CLASS}) for the objects to train and test our methods on. It contains $1.2 \times 10^6$ objects to a limiting magnitude of $F814W = 26.5$ from images observed with the Hubble Space Telescope (HST) using the Advanced Camera for Surveys (ACS)\NS{\footnote{\url{http://www.stsci.edu/hst/acs/documents/handbooks/current/cover.htm}}}, therefore its image quality (very deep and unaffected by the atmosphere) can be used as a `truth' reference. Images were taken through the wide F814W filter ($I$). The catalog contains, roughly, $1.1 \times 10^6$ galaxies, most towards the faint end, 30,000 stars and the rest are fake detections\footnote{Technically, this classification only separates point-like versus extended objects. \NS{QSOs will tend to be mixed with both samples and are neglected in this work, as they are estimated to be $\sim 3\%$ in the COSMOS field. This can however be a very interesting avenue to explore in the context of \cnn narrow band classification.}}.

\subsection{PAUS}
The Physics of the Accelerating Universe Survey was born in 2008 with the idea of measuring precise redshifts\footnote{$\approx 0.35 \%$ error, meaning a precision of $\sigma (z)$/(1+$z$) $\approx$ 0.0035, versus a typical $5\%$ for broad band measurements} for a large number of galaxies using photometric measurements \citep{martipau}.  The novelty of the project was to carry out a photometric survey with a large number of narrow filters, in order to obtain a low-resolution spectrum of a large number of cosmological objects. Among other science cases, this will allow the study of clustering at intermediate scales \citep{Stothert}, intrinsic alignments \NS{of galaxies, that can bias cosmological measurements using weak lensing \citep{desy1cosmo}} or contributing to the effective modeling of galaxies in image simulations \citep{tortorelli}.\\

The PAUS camera, named PAUCam  \citep{padilla}, is equipped with 40 narrow band (NB) filters, 13 nm wide and separated by 10 nm, covering a total wavelength range from 450 nm to 850 nm and 6 \textit{ugrizY} broad band filters (which are not used in this work). The survey covers approximately  0.75 square degrees of equivalent 40 NB area per night, delivering low-resolution $(R \approx 50)$ spectra for all objects in the field of view. The camera is mounted at the prime focus of the 4.2m William Herschel Telescope. As of May 2018, PAUS has been observing for  approximately 40 nights per year since mid-2015 (with an efficiency below 50\% due to bad weather).\\  

PAUS data is managed by a complex infrastructure which starts at the mountaintop, stores data temporarily there and sends it to the PAUS data center at the Port d'Informaci\'o Cient\'ifica (PIC) where the nightly and higher level pipelines (Serrano et al. in prep.) are run and data is archived for long term storage, as well as distributed through a database for scientific use (\citet{carretero}, Tonello et al. in prep.). \\

Photometric calibration is tied to the Sloan Digital Sky Survey \citep[SDSS,][]{sdss_standards} stellar photometry. Each PAUS image is separately calibrated using high signal-to-noise detected stars that are matched to the SDSS catalogues. The SDSS broad band photometry for these stars is fit to the Pickles stellar templates \citep{pickles}\footnote{\url{http://www.stsci.edu/hst/observatory/crds/pickles_atlas.html}} to obtain a spectral energy distribution (SED), which is used to synthesize the expected NB fluxes in PAUS. Single image zero points are then determined by comparing the modeled and observed fluxes (see Castander et al. in prep. for more information). \LC{The photometric error is estimated to be $\sim 1-2\%$. It is slightly above $1\%$ for the redder bands, and increasing towards the bluer end.}  \\

The PAUS catalog over this field contains 49,000 astronomical objects, matched to the COSMOS catalog and with the 40 NB measured: 42,000 galaxies and 7,000 stars from magnitudes $I=16$ to $I=23$. \LC{The flux units of such objects are given in electrons per second}. \\ 

\LC{The training set size for PAUS objects with $I < 22.5$ is 20,000 objects, from which 15,000 are galaxies and 5,000 are stars. For validation, the sample size depends on the magnitude range we are testing. For objects with $I < 22.5$, we use 6,000 objects, 1,000 stars and 5,000 galaxies, decreasing for smaller magnitude ranges. It is worthwhile noting that a small percentage of the training set will include QSOs labeled as 'stars' in our case (around $3\%$ of the total stellar sample in COSMOS, according to the milliQUAS compilation\footnote{\url{http://quasars.org/milliquas.htm}}\citep{flesch}), so a further optimization could be possible by identifying these.\\\COMMENT{However, the validation sample size should not affect the results of the network, as machine learning algorithms are training size dependent, but not validation size dependent.}}

\subsection{ALHAMBRA}
We have also used the ALHAMBRA photometric redshifts catalog \citep{Molino} over the ALHAMBRA-4 field, which overlaps with COSMOS. It contains 37,000 objects matched to our reference COSMOS catalog, from which 34,000 are galaxies and 3,000 are stars. The ALHAMBRA photometric system \citep{Aparicio-Villegas} is characterized by 20 constant width (31 nm), non-overlapping medium band filters covering a wavelength range from 350 nm to 970 nm. The images were taken using the Calar Alto 3.5m telescope using the wide field optical camera LAICA and the NIR instrument Omega-2000, which are equipped with 20 intermediate width bands and 3 NIR broad bands: $J, H, K$.The catalog presents multicolor PSF-corrected photometry detected in synthetic F814W images with objects up to a magnitude of $F814W \approx 26.5$.\\

\LC{The catalog we will work with contains only objects with less than 5 undetected bands. It contains 29,000 galaxies and 2,700 stars.}. From this catalog, those with magnitude brighter than 22.5 are 7,600: 5,900 galaxies and 1,700 stars. \LC{For these objects, the training size we are using contains 1,500 stars and 5,000 galaxies. However, as ALHAMBRA goes deeper than PAUS, we could also train with objects up to magnitude 26. For these objects, the training contains 2,300 stars and 15,000 galaxies.  In the ALHAMBRA catalog whenever a source was not detected in a given band, its magnitude was set to a 'sentinel' value of 99. }

\section{Methods}
\label{sec:method}
Artificial Neural Networks, \LC{Random Forests and Convolutional Neural Networks} have already been used to classify stars and galaxies successfully \citep[as shown, e.g., in ][]{Soumagnac,Kim15,Kim}, however, as mentioned before, they have never been used solely with \LC{ photometric measurements of the objects spectra, without additional morphological information.}\COMMENT{band fluxes inputs.} \COMMENT{On the other hand,} Convolutional Neural Networks (\cnn) have \LC{also} been applied to different fields with excellent results, for instance in medical imaging \citep{Qayyum}, and they have proven to be very powerful in image processing and pattern recognition, also \NS{for the case of} one dimensional information \citep{mendez}\NS{, where levels of radon in the environment can very successfully be predicted using CNNs learning from the shape of fluctuations of previous behaviour}. These algorithms have also been applied to spectral classification \citep{Pavel} and to tackle the star-galaxy classification problem using whole CCD images as input feature map \citep{Kim}. 

\subsection{Machine Learning algorithms}
In this section, we describe the three machine learning algorithms for which we have compared  performances in our case of study. 
\subsubsection{Artificial Neural Networks (ANN)}
Neural networks \citep{werbos} are a biologically-inspired programming paradigm that enables a computer to learn from observed data. They can be applied to difficult classification tasks, where a training sample already classified by other means is used to 'teach' the network. The learning process consists in recursively weighting the input features (the fluxes on the different bands in our case) by some factors, the weights, chosen in order to optimize the classification algorithm. This consists \LC{of} the evaluation of a 'cost function', which is a measure of the overall agreement between the actual nature of the objects in the training sample and that inferred from the weighted inputs.\\

An \nn consists of a set of layers, an input layer and an output layer at the beginning and end of the network and a set of hidden layers in the middle, each of them containing a set of trainable weights. The goal of the network is to optimise the set of weights to those that minimise the error in the network prediction.

The network is provided with a loss function that estimates the agreement between the prediction of the network and the truth  value. The loss function is evaluated after every iteration on data and the loss value obtained is used to back-propagate the network. Back-propagation is nothing but an optimisation technique that modifies the set of weights of the neural network in order to minimise the total loss function.
It consists in the loss differentiation with respect to the weights, so that it is estimated how much a change in a given weight affects the total error. These derivatives are the gradients of the loss function with respect to the weights. The gradients are used to update the weights to those that will be used in the following data iteration. Back-propagation starts in the output layer and goes through all the layers of the network until it arrives at the input layer, updating all the weights, which is why it is called back-propagation. When the cost function can no longer be minimised by a substantial amount, the weights are saved and the resulting model is used for classification.

An iteration over the full training sample is called epoch. However, it is common practice to divide such training sample in batches and propagate the network over each of these batches instead of using all the training sample simultaneously. This allows the network to update itself more frequently. The size of these batches is called batch\_size and it is an important network parameter.\\

We have used the neural network implementation from the Python \texttt{scikit-learn} package \citep{scikit-learn}. \LC{The neural network architecture used consists of an input and an output layer with three intermediate hidden layers, with 40 hidden neurons (weights) per layer.}

\COMMENT{The combination of weights which minimises the cost function is considered to be a solution of the learning problem.} 

\subsubsection{Random Forests (RF)}
\COMMENT{As with neural networks, a random forest algorithm is also a supervised classification algorithm} \LC{A Random forest}, \citep{breiman} is composed of a collection of \COMMENT{simple} decision tree predictors, each of them giving an output class when given a set of input features. \\

A decision tree classifies data items executing step-by-step choices by posing a series of questions about the features associated with the objects.  Each question is contained in a node and each node leads to children nodes, one per possible answer to the parents' node question. \LC{Each question splits the data as it progresses through the algorithm,  \COMMENT{The questions therefore} forming a hierarchy encoded as a tree.} The training set is used to establish the features' hierarchy and the value of the \COMMENT{cuts} \LC{splits} in each of the nodes which optimises the classification. After each iteration over the whole tree, the separation power is evaluated and the \COMMENT{cuts} \LC{splits} are selected according to it.\\

In a Random Forest approach, many different decision trees are created. The training set is sampled with replacement so as to produce a training set for each of the decision trees taking part of the forest. \LC{The term 'with replacement' means that, when a given object is sampled for a given decision tree dataset, such object is not removed from the complete dataset, in such a way that different decision trees may share objects in their datasets.} Another difference is in the choice of the question at each node. In the random forest approach, only a random subset of the features is considered. Therefore, each decision tree shaping the forest may give a different classification output for the same sample. The prediction output is a combination of all the particular results by taking the most common prediction.\\
As with the case with neural networks, we have also used a random forest implementation provided by the \texttt{scikit-learn} Python package. \LC{The architecture of our random forest consists of 35 decision trees, with the maximum number of features to consider when looking for the best split equal to the square root of the total number of features and without a predefined maximum depth of each tree, letting the trees expand until each leaf is as pure as possible.}

\subsubsection{Convolutional Neural Networks (\cnn)}
Convolutional Neural Networks \NS{\citep{lecun89,lecun98}} are a category of neural networks that have proven very effective in areas such as image recognition and classification. One characteristic of \cnn compared to its predecessors is its ability to recognize patterns based on local features. \\

\LC{One can find different types of layers in a \cnn, each of them applying different operations on the data. Particularly, our network is built with three different layers: the convolutional layer, the pooling layer and the fully connected (or dense) layer. As with  artificial neural networks, the input data is propagated through the different layers of the network in batches of a fixed size.\\
Convolutional layers convolve the input data with a weight matrix,  named kernel, which contain the learning parameters. The convolution consists of the multiplication of the kernel by the input map, where the kernel size and the stride between consecutive convolutions are fixed.  The output of this layer is a set of feature maps resulting from the convolution of the initial input. A remarkable aspect of \cnn is its local connectivity. The layers in a ANN are fully connected, which means that all neurons from a layer are connected to those on the layer below. Conversely, CNNs are locally connected: each neuron only receives input from a small local group of the pixels in the input image, which coincides with the user's defined kernel size. The aim of locally connected layers is to allow for the detection of some subtle nuances of spatial arrangements which are common to the specific spectra we are classifying, independently of their position in wavelength.}\\

\LC{Another type of layers are pooling layers.} They are used to reduce the dimensionality of the feature maps. There are different pooling methods carried out by different functions across local regions of the input. One usual pooling function is the maximum, which consists in grouping features together and keeping only that containing the largest value, although another typical alternative uses the mean function instead. This layer reduces the number of operations required for all the following layers while still passing on the valid information from the previous layer. \COMMENT{The trainable parameters are located in the convolutional layer and typically, the pooling layer does not contain any.} \LC{The pooling matrix size is also a user's defined parameter, as well as the stride between poolings.} \\

The final \cnn output is generated through a fully connected layer (also called dense layer). It applies a linear operation in which every input is connected to every output by the weight to generate an output with dimensionality equal to the number of output classes we need. \LC{In fact, these layers are the \nn linear layers.} The output layer contains again a cost function that  evaluates the error in the prediction. Similar to the \nn, once the forward pass is complete the back-propagation begins to update the weights for loss reduction.\\

\begin{figure*}
	\centering
    \includegraphics[width=1.0\textwidth]{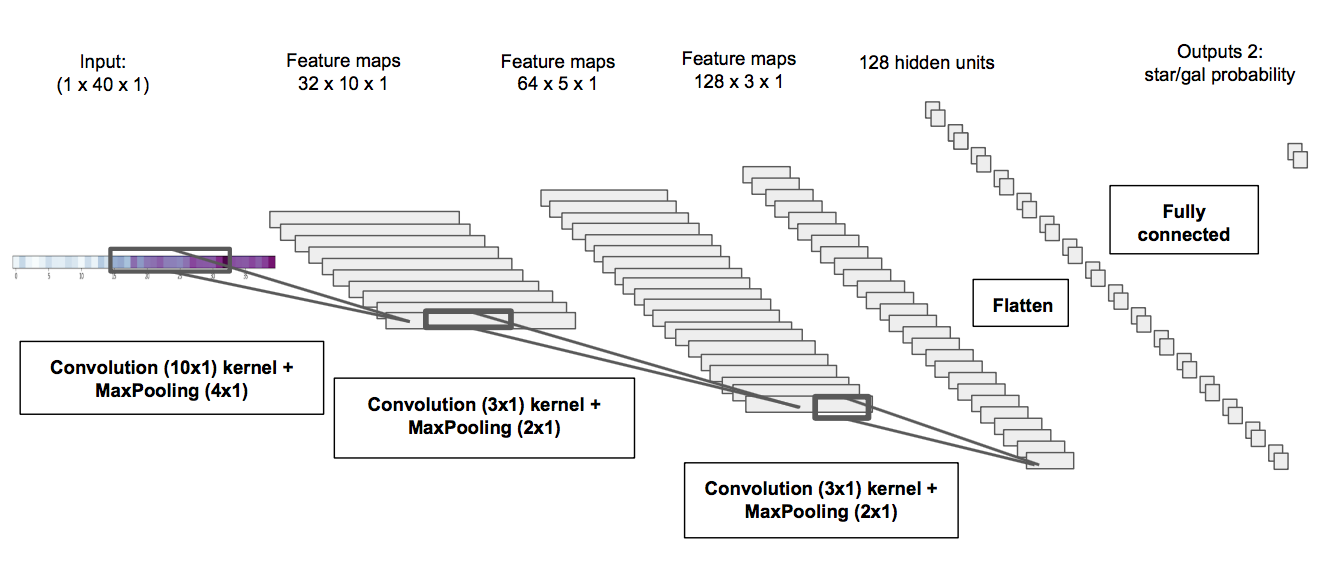}
    \caption{The \cnn architecture used for this paper. The input data is a (40x1) dimensional array containing the flux in the 40 PAUS' narrow band photometric filters. The output is the probability of being a star or a galaxy.}. 
    \label{theo_convNet}
\end{figure*}

\LC{Figure \ref{theo_convNet} shows the architecture of our \cnn. The input of the network is a (40x1) dimensional array containing the 40 PAUS photometric fluxes. The network contains three convolutional layers with kernel sizes 10,3,3 respectively. Such convolutional layers are provided with an activation function, which in our case is a `LeakyReLu' function (from \texttt{Keras}, see below). The first convolution is larger, so that the algorithm learns about more general features. The following convolutional layers have a smaller kernel to focus on more subtle nuances. Each convolution is followed by a pooling layer. The pooling sizes are 4,2,2 respectively. After the last pooling layer, the dimensions of its output are converted into a flat array, an array collapsed into one dimension, in what is called 'Flatten'. The output flattened array is the input of the fully connected layer, which connects directly to the output layer of the network. The fully connected layer and the output layer have dimensions 128 and 2, respectively, corresponding to the dimensions of the last pooling output, 128, and the output of the network, 2 (star and galaxy classes). The final output corresponds to the object's probability of being a star and that of being a galaxy (technically, however, both add up to unity in our case). We have used the \texttt{Keras} Python library \citep{keras} to build our algorithm.} \\ 

\LC{It is also worth to mention that it is not possible to train the \cnn with missing bands: a \cnn algorithm cannot be trained if the input contains gaps in some of the bands. There are algorithms that can fill these missing bands with different methods (mean value of the whole input, nearest neighbors, etc.). In case of need, we would fill the gaps with linearly interpolated features based on its contiguous neighbors. In order to have reliable measurements to test our method, we would only keep objects with 5 or less non-detected bands. We will use this method in the ALHAMBRA section, so that we have a larger dataset to train with.}

\subsection{Analysis}
To analyze the performance of the classifiers, we will often refer to \textit{precision} or \textit{recall}, and receiver operating characteristic curves, \textit{ROC} curves. In the context of this paper, a positive result means an object classified as a galaxy whereas a negative result refers to any object classified as a star. With such terminology, Table \ref{ROC_definitions} defines the concepts of true and false positive and true and false negative contextualized to our problem. 
\begin{table}
\centering
\caption{TP stands for 'True positive', FP for 'False positive', FN for 'False negative' and TN for 'True negative', for a given threshold.}
\label{ROC_definitions}
\begin{tabular}{|c|c|c|}
\hline
\textbf{}            & \textbf{Classified galaxy} & \multicolumn{1}{l|}{\textbf{Classified star}} \\ \hline
\textbf{True Galaxy} &  TP                         &  FN                                           \\ \hline
\textbf{True star}   &  FP                      &  TN                                          \\ \hline
\end{tabular}
\end{table}
With such parameters, we can define the True Positive Rate (TPR) and the False Positive Rate (FPR) (Equations \eqref{TPR} and \eqref{FPR}) and also, the precision and the recall (Equations \eqref{precision}.\COMMENT{, \eqref{completeness}).}

\begin{equation}
TPR = \frac{TP}{TP+FN} = Recall,
\label{TPR}
\end{equation}

\begin{equation}
FPR = \frac{FP}{FP+TN},
\label{FPR}
\end{equation}

\begin{equation}
Precision = \frac{TP}{TP+FP},
\label{precision}
\end{equation}
\\

\COMMENT{
\begin{equation}
Recall = \frac{TP}{TP+FN} = TPR,
\label{completeness}
\end{equation}}

The performance of the classifiers is generally studied in the ROC space by ROC curves. A ROC curve is a graphical plot that illustrates the diagnostic ability of a binary classifier system as its discrimination threshold is varied (the limit on a given classifier for which an objects is considered to belong either to a class or another), using the True Positive Rate vs False Positive Rate values typically. The area under the curve (AUC) gives a measurement of the performance of the classifier, where an area of 1.0 would mean a perfect classifier. A diagonal through the plot would indicate a random performance (therefore with an AUC $\sim 0.5$).\\

For our case of study, the algorithms output is the object's probability of being a galaxy. The ROC curve shows the True Positive Rate (the number of galaxies classified as galaxies over the total number of galaxies) against the False Positive Rate, (the number of stars classified as galaxies over the total number of objects classified as galaxies) when the probability threshold for which an object is considered either a star or a galaxy is varied.
The ROC curve could also be represented with the True Negative Rate and the False Negative Rate, rating the classification/misclassification of stars instead of galaxies. 

\section{Algorithm performance}
\label{sec:characterization}
In this section, we analyze concurrently the performance of the three algorithms defined in section \ref{sec:method}: artificial neural networks, random forests and convolutional neural networks.\\  We use a training sample over the COSMOS catalog, matched to PAUS objects, where their 40 narrow-band fluxes have been used as the input data vector, up to magnitude $I=22.5$ as defined by our reference COSMOS catalog. 

\subsection{Training set size dependence}
\label{sec:imbalanced}
The performance of any machine learning algorithm is related to the number of samples used in the training phase. However, using too many training samples may be self-defeating; the training could take much longer than \LC{required}. \\

\begin{figure*}
     \begin{center}
        \subfigure[ROC-AUC values for different training sizes.]{%
            \label{AUC_size}
            \includegraphics[width=0.45\textwidth]{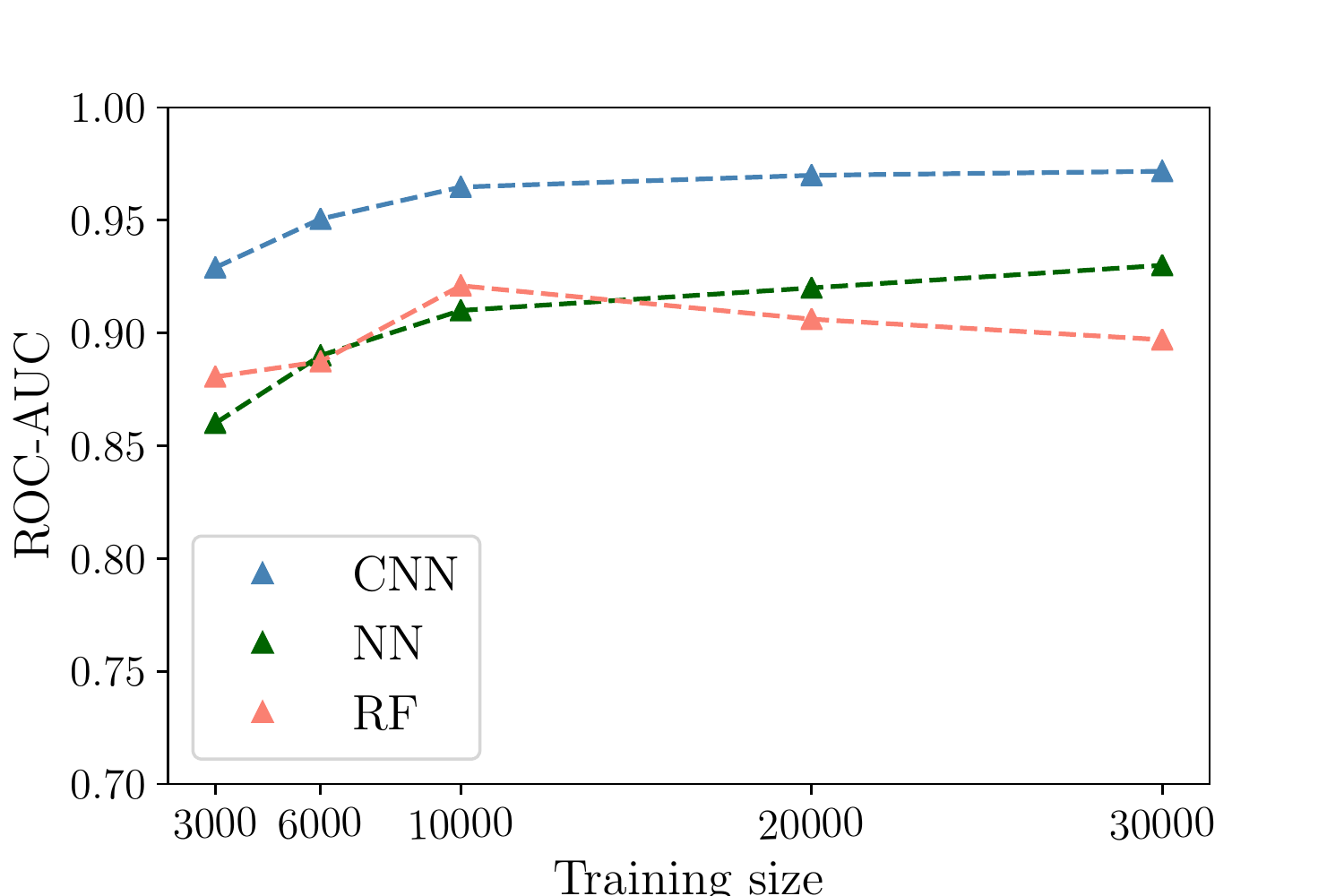}
        }
        \subfigure[ROC curves corresponding to different number of training objects for the \cnn.]{%
           \label{ROC_size}
           \includegraphics[width=0.45\textwidth]{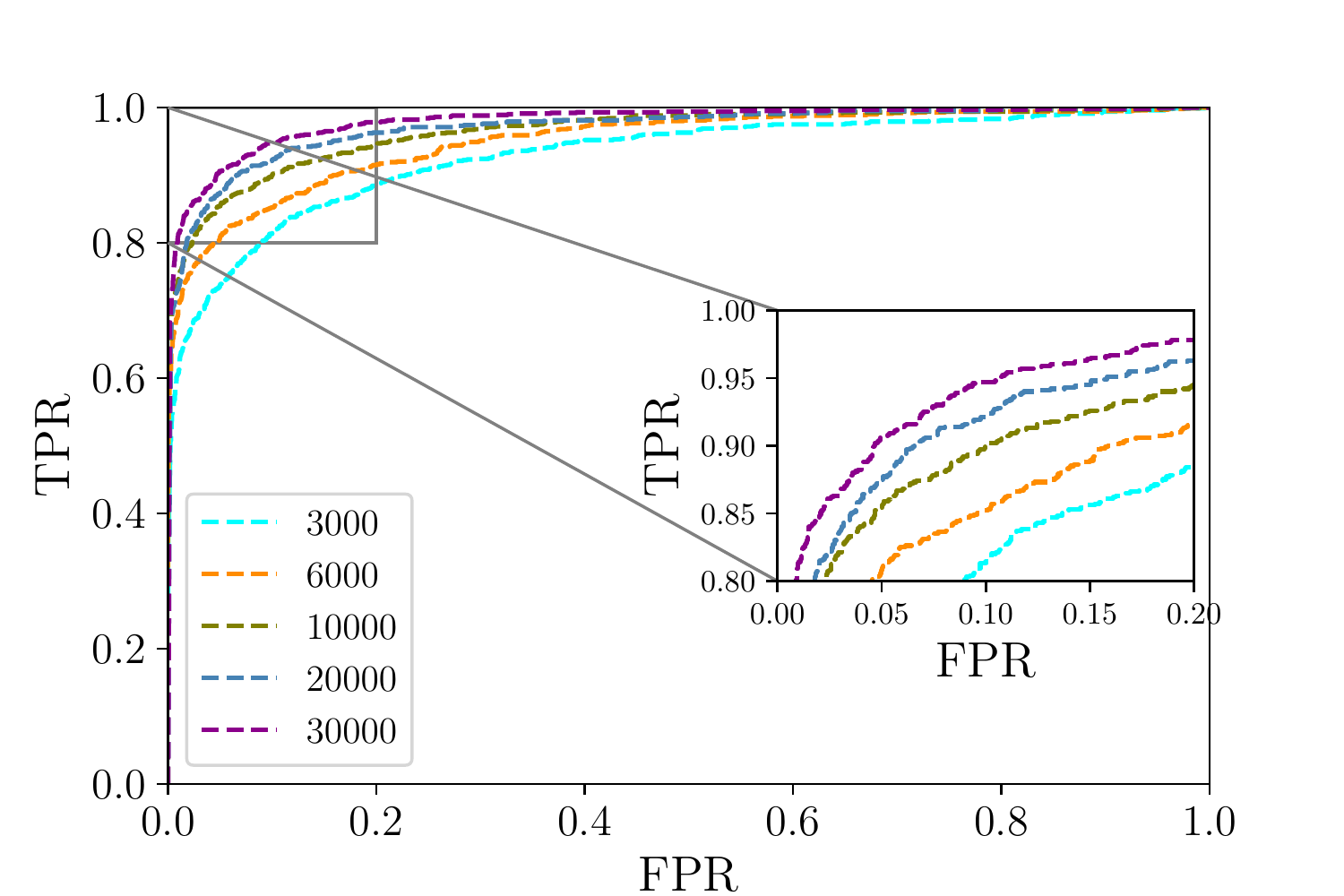}
        }\\       
    \end{center}
    \caption{%
        (a) ROC-AUC scaling for different training sample sizes for the \nn, random forests and the \cnn using PAUS data with $i_{auto} < 22.5$ and 40 NB inputs. The results plotted are those obtained on the validation sample. (b) ROC curve showing the scaling of the \cnn performance with the number of training objects}  %
   \label{training_size}   
\end{figure*}

Figure \ref{AUC_size} shows the training size dependence for the neural network, the random forest and the convolutional neural network, where the results plotted are those obtained on a distinct validation sample. \LC{We only vary the size of the training sample, while maintaining the validation sample size constant.} \\

One can already see here that the \cnn is the algorithm yielding the best classification, with a better performance than those of the random forest or the \nn (see section \ref{sec:algcomp} for discussion). It also showcases that for this sample, 10,000 objects are enough to get high classification rates with the \cnn. Nevertheless, Figure \ref{ROC_size} shows that from 10,000 to 30,000 objects the classification is still improving, although the improvement is smaller than from 3,000 to 10,000 objects. This means that the algorithm is more sensitive to training sample size increments when the training datasets are small.\\

\subsection{Number of input bands dependency}

In any classification problem, the more information is available about each class, the easier it is to identify particular patterns useful to differentiate between them. For the case of astronomical object classification, any spectral related or morphological information may be meaningful. However, a large number of input features can also have its drawbacks. Over-fitting or scaling problems may arise from such a large dimensionality. When the input's dimension increases, the hypervolume in input feature space increases so fast that the available data becomes sparse. Also, the data needed to provide reliable results increases exponentially.\\

For PAUS, the 40 available optical bands are probably not enough to encounter such problems. However, it is also of interest to check this and study how the performance scales with the number of bands (i.e., with the spectral resolution). To see how the algorithm scales with the loss of resolution, we have merged the 40 PAU NB in groups of 2,4,5 and 8 summing the flux of contiguous bands and therefore providing data samples with 20,10,8 and 5 bands, respectively. \\

\begin{figure*}
     \begin{center}
        \subfigure[ROC-AUC values for different number of inputs bands.]{%
            \label{AUC_nfilters}
            \includegraphics[width=0.45\textwidth]{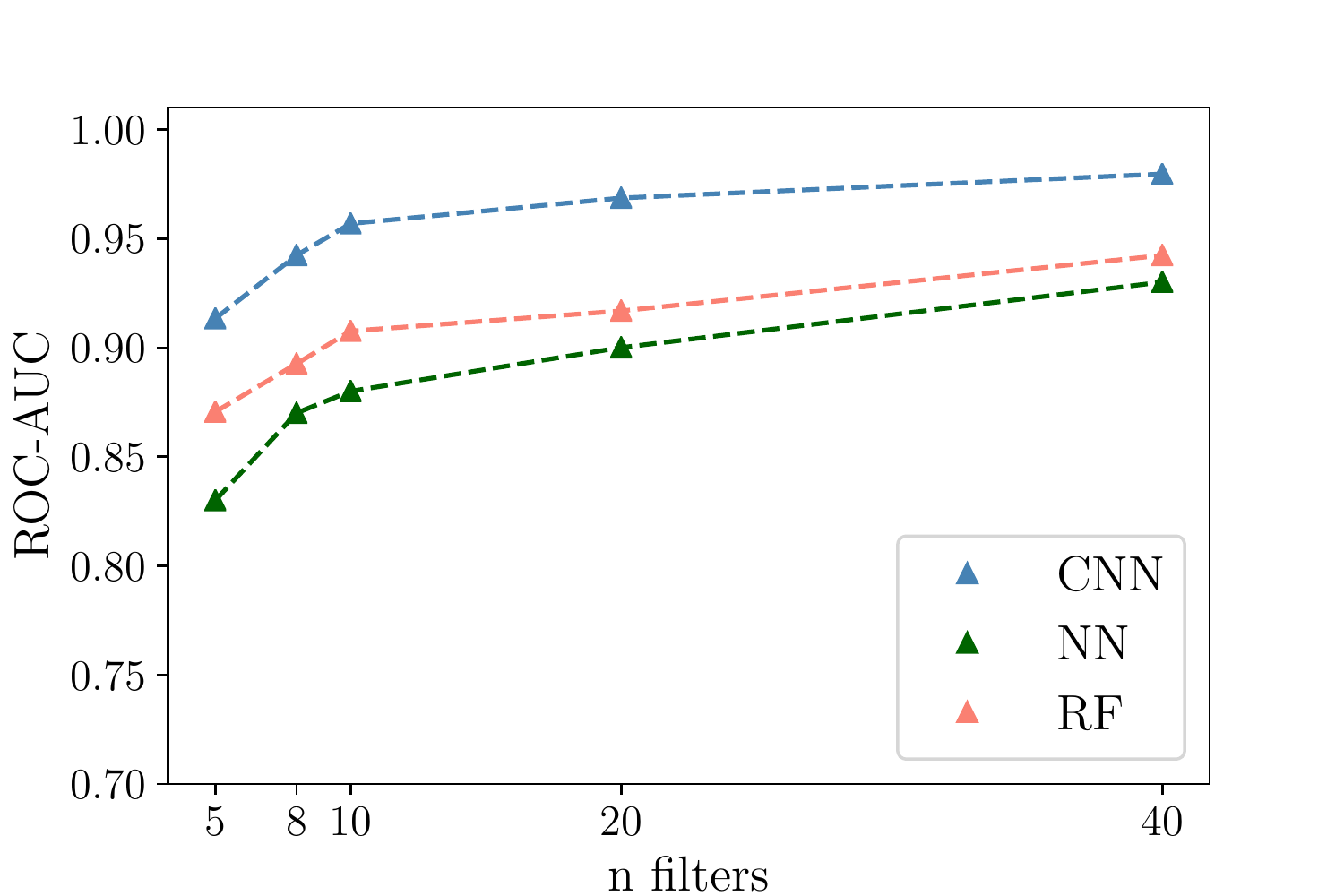}
        }
        \subfigure[Algorithm performance dependency on input number of bands.]{%
           \label{ROC_nfilters}
           \includegraphics[width=0.45\textwidth]{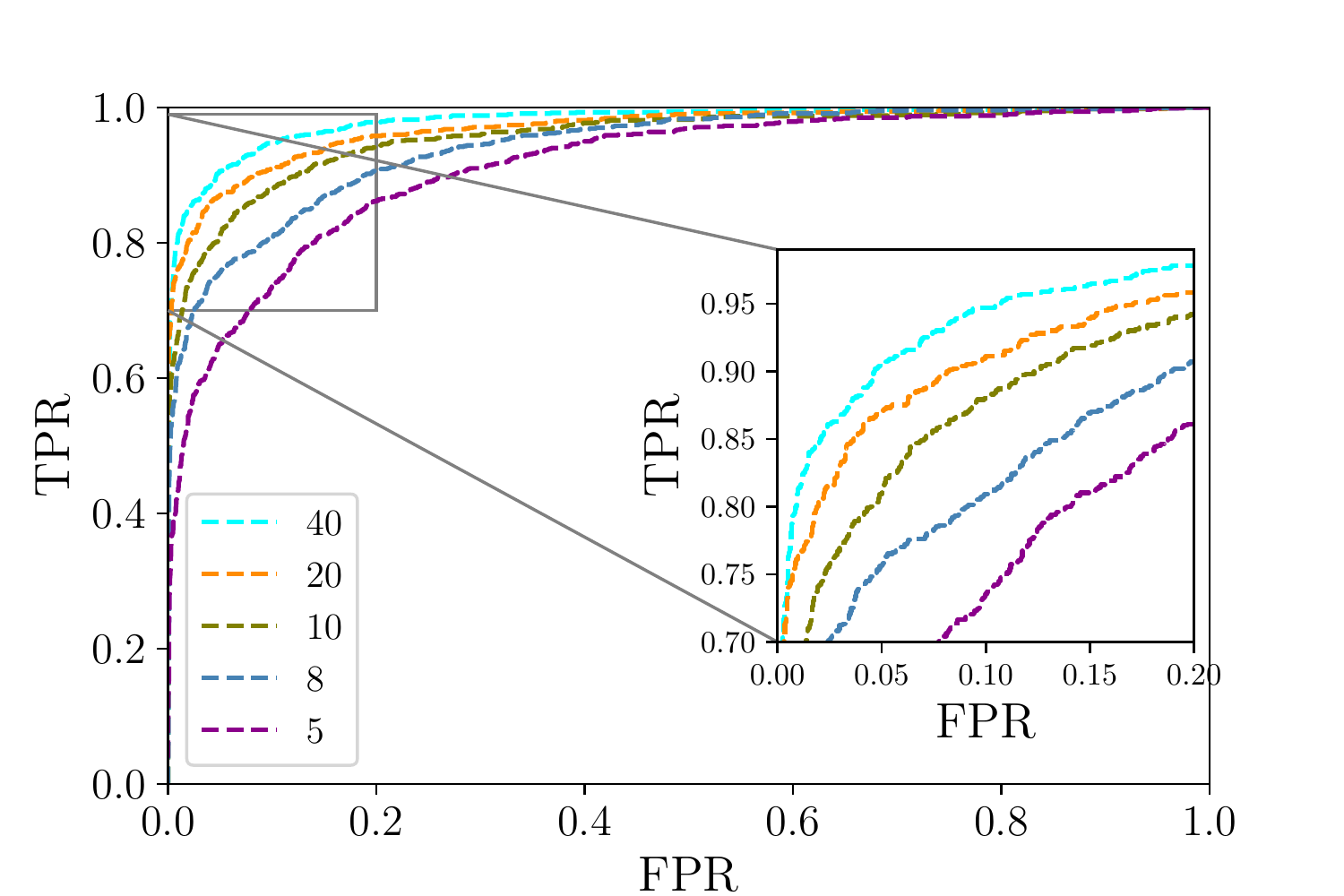}
        }\\       
    \end{center}
    \caption{%
         (a) ROC-AUC scaling for different number of bands for the \nn, the RF and the \cnn using PAUS data with $i_{auto}<22.5$ and 10,000 training objects. The results plotted are those obtained on the validation sample. (b) ROC curves showing the scaling with the spectral resolution with the \cnn algorithm.}%
   \label{AUC_bands}   
\end{figure*}

Figure \ref{AUC_nfilters} presents the scaling with the number of bands for the three different algorithms, exhibiting the same pattern in all cases: as the spectral resolution increases, the performance of the algorithms improves. Such improvement is not linear; it has a more significant slope from 5 to 10 bands in all cases.  One can see that with the \cnn the photometry does not need to have 40 narrow bands to already give a good classification of stars and galaxies: 20 bands already result into high classification rates, as the shapes from the spectrum used to differentiate these two cases are already evident at such resolutions. \\

Figure \ref{ROC_nfilters} shows the different performances in the ROC space for the \cnn. It exhibits the results we have already mentioned: there are important gaps between the curves from 5 to 20 bands, whereas increasing from 20 to 40 bands translates into a smaller improvement.\\

We have also studied the difference between using the 20 bluest bands versus the 20 reddest. The ROC-AUC for the bluer bands is 0.913 $\pm$ 0.005 whereas for those redder bands, it is 0.950$\pm$ 0.004. Therefore, we find that star-galaxy separation is therefore more sensitive to the information contained in the redder bands \NS{in the wavelength range of PAUS}, as many of the stars are typically red dwarfs with characteristic absorption features.
There is also another effect one could consider: the bluer bands have lower S/N than the redder ones and one would expect that the classification performs worse.

\subsection{Algorithm comparison}
\label{sec:algcomp}
We have shown that the \cnn is  exhibiting the best performance so henceforth it will be the fiducial algorithm applied to the classification on the PAUS and ALHAMBRA catalogs.\\

\begin{figure*}
     \begin{center}
        \subfigure[Galaxies.]{%
            \label{}
            \includegraphics[width=0.45\textwidth]{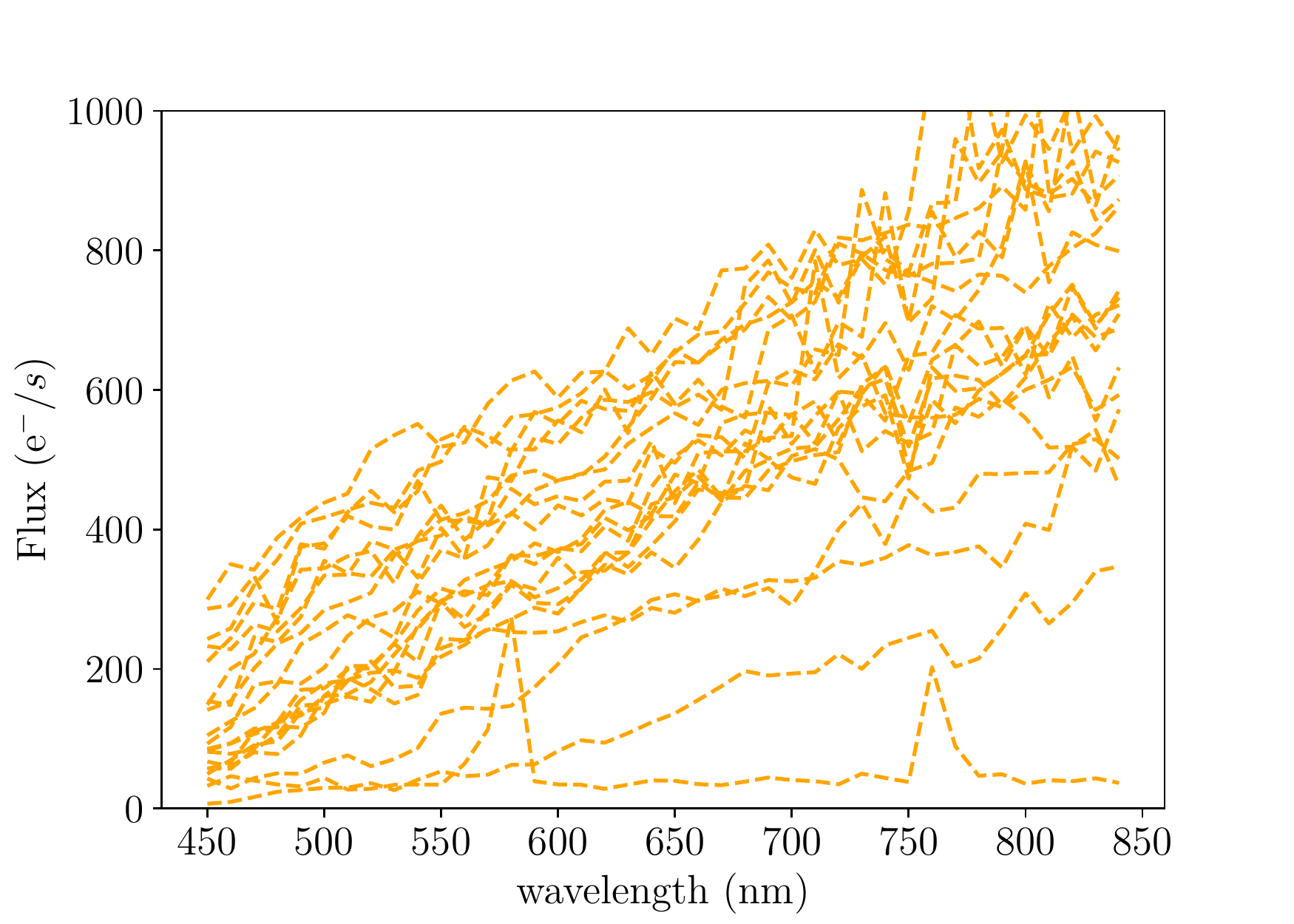}
        }
        \subfigure[Stars.]{%
           \label{}
           \includegraphics[width=0.45\textwidth]{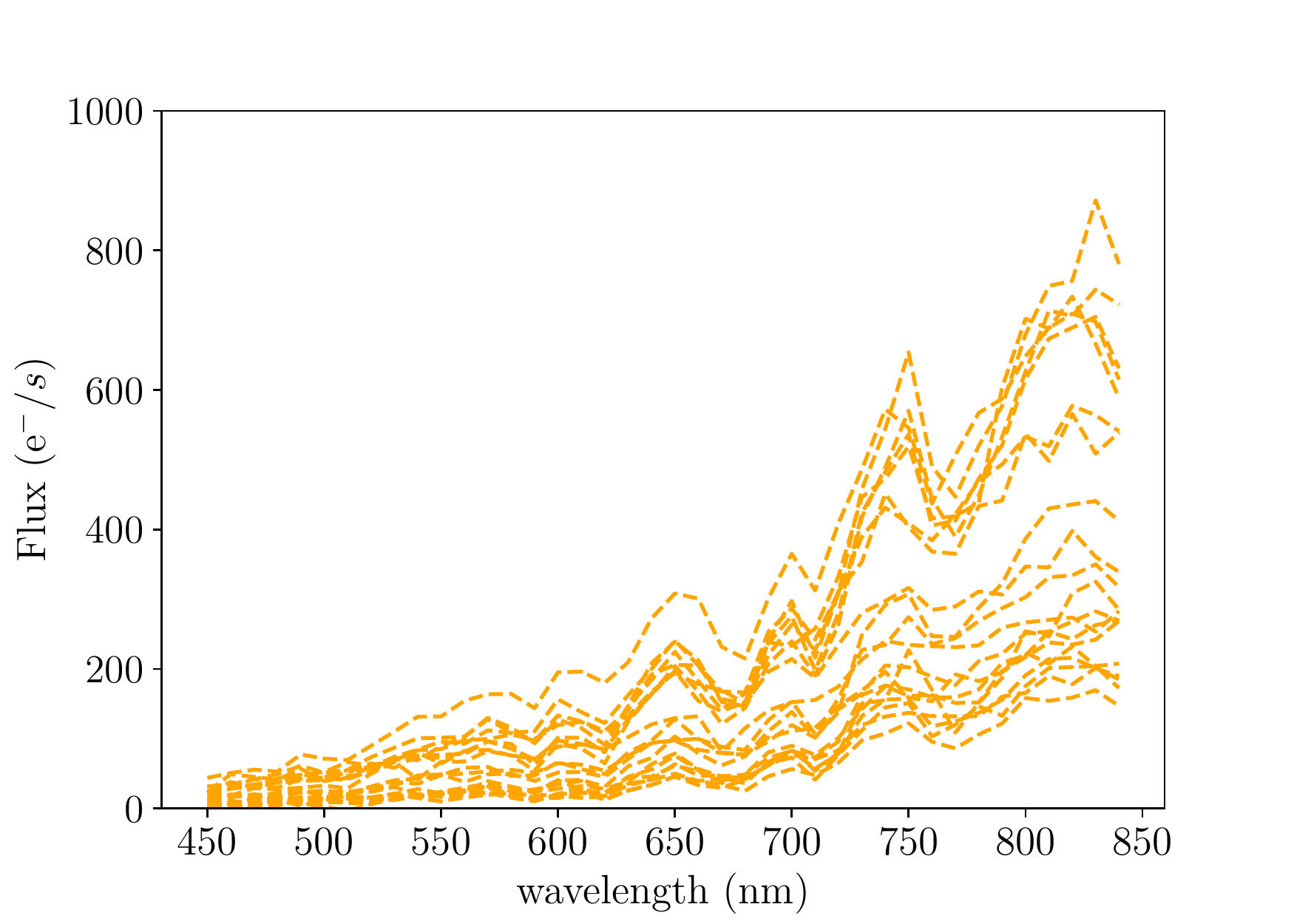}
        }\\       
    \end{center}
    \caption{%
         (a) Spectra of objects classified as galaxies with the CNN with probability $\sim1$. (b) Same for stars.}%
   \label{objecs_patterns}   
\end{figure*}

There are many effects that are contributing to this result. As was mentioned above, \cnn are provided with locally connected layers that are capable of recognizing subtle nuances of spatial arrangements. Figure \ref{objecs_patterns} shows objects classified as galaxies (a) or stars (b) with high probability by the \cnn. In the case of galaxies, one can notice that many of them contain \LC{peaks} in one or two consecutive bands that most likely correspond to emission lines, meaning that the \cnn is able to learn from these characteristic traits. We have made the same check with the \nn or the random forest algorithms and none of them present clear emission line patterns in the best classified galaxies. For stars, one can see that there are many objects with the same spectral shape, including certain patterns (peaks, valleys) usually in approximately nearby sections of the spectrum. Most of these correspond to red stars which in general are more commonplace in the dataset at the considered magnitudes. The algorithm in these cases is able to recognize these objects from the training set so that they can be identified readily as stars.
This result (good performance of \cnn on 1D quasi-spectral data for classification) is a true finding of this work, which opens up possibilities only available to this kind of photometric surveys, in which object types could be classified for large sets of objects without previous selection \NS{(as opposed to what is done in spectroscopic surveys)}.

\section{Results}
\LC{Previous sections showed that the \cnn is the algorithm giving better performances in the star/galaxy classification using PAUS multi narrow band photometry. In what follows, we will present the results of the application of this algorithm to the PAUS and ALHAMBRA datasets.}
\label{sec:results}
\subsection{Classification on the PAUS catalog}
As explained in section \ref{sec:data}, the PAUS catalog in the COSMOS region contains 49,000 objects up to magnitude $I=23$, from which 7,000 are stars and 42,000 are galaxies. We will work with a subsample of objects with $I < 22.5$, for which the catalog contains 6,000 stars and 28,000 galaxies. \LC{As stated in section \ref{sec:data}}, the training sample employed to carry out the classification is composed of 20,000 objects up to magnitude $I=22.5$, 15,000 galaxies and 5,000 stars, whereas the validation sample contains 1,000 stars and 5,000 galaxies.
In section \ref{sec:characterization}, we already noted that \cnn  is the best choice for classifying stars and galaxies using band fluxes input, and therefore we will use it by default in the rest of this work.\\

The PAUS catalog contains objects with negative flux measurements. This may happen for sources with a very low signal-to-noise in a given band and for which the background has been overestimated. In these cases, it is not possible to estimate a magnitude, and the corresponding value in the catalog is set to a 'sentinel' value of 99.0. However, in this section we will use the PAUS fluxes as inputs for the algorithm and therefore, the negative counts do not translate into a problem when training the network.\\

The algorithm outputs a probability of the object being either a star or a galaxy, which we will call \cnnc. The resulting ROC-AUC is $0.973 \pm 0.001$, leading to a purity of \NS{$98.4\pm0.1\%$} for a completeness of \NS{$98.8\pm0.1\%$} for objects brighter than $I=22.5$. This means that, the selected galaxy sample still contains a \NS{$1.6\%$} of stars contaminating it, while losing a \NS{$1.2\%$} of the original true galaxies of the sample.\\

\begin{figure}
	\centering
    \includegraphics[width=0.5\textwidth]{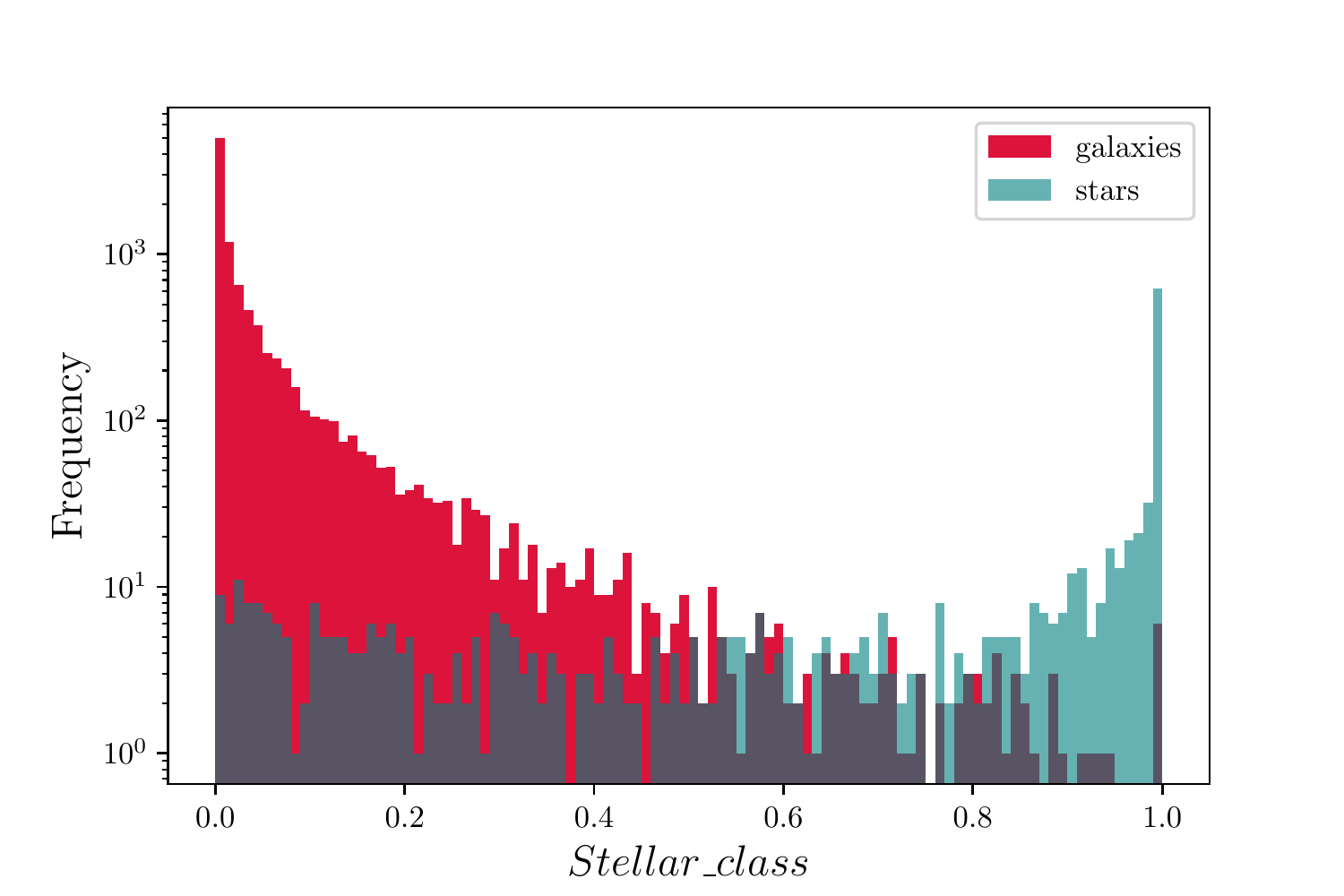}
    \caption{Distribution of $cnn\_stellarity$ for stars (blue) and galaxies(red), both populations on the validation sample. Darker regions correspond to the overlapping of stars and galaxies.}%
    \label{Stellar_class_Hist}
\end{figure}

\begin{figure}
     \begin{center}         \includegraphics[width=0.5\textwidth]{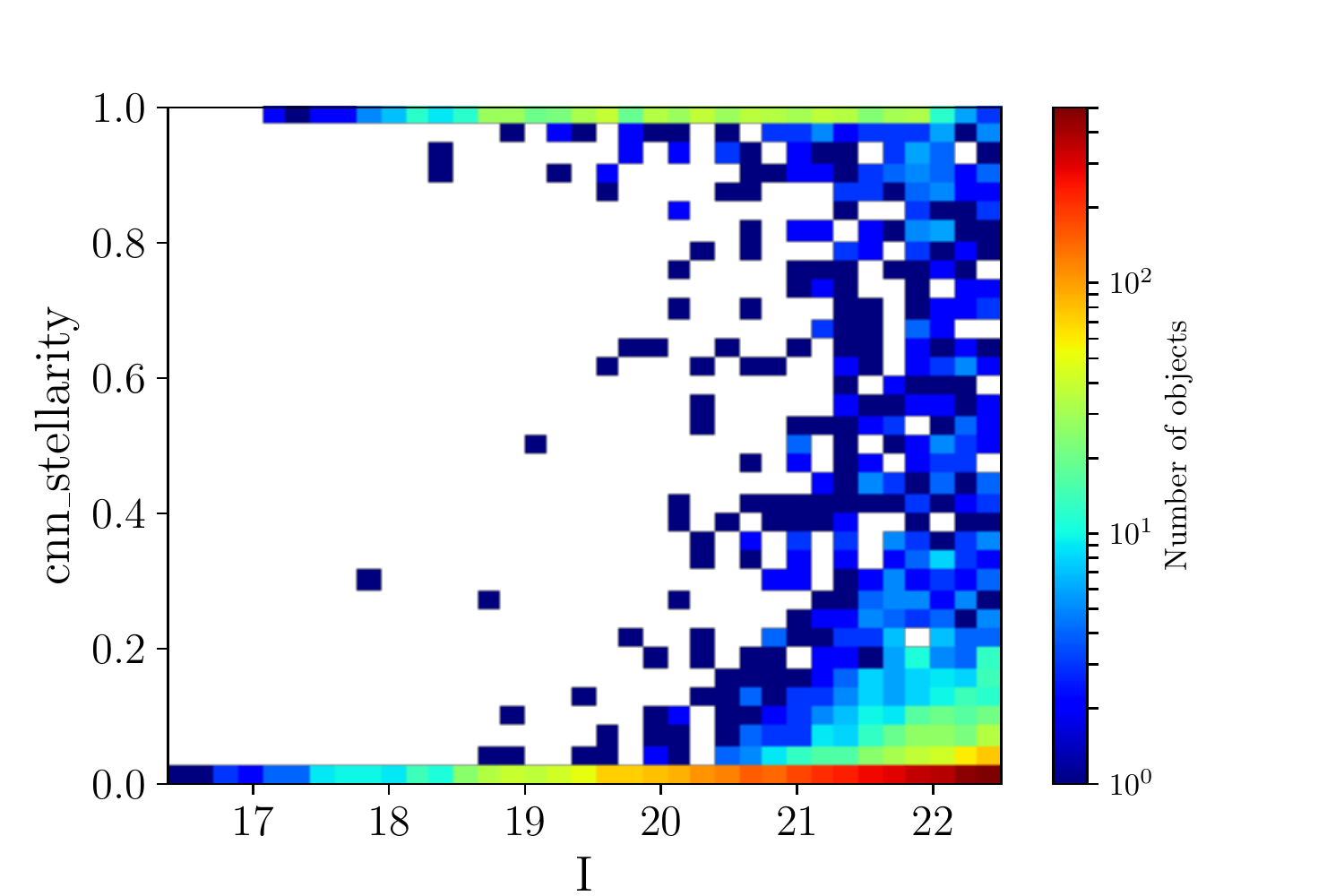}
      \end{center}
    \caption{Heatmap for \cnnc as as function of $I$ magnitude, as measured by the HST-ACS on the COSMOS field.}%
   \label{heat-map}   
\end{figure}

Analyzing in more depth the classification of the PAUS sample, Figure \ref{Stellar_class_Hist} shows the histogram of such output probability. It exhibits two clearly differentiated peaks in 0 and 1, which correspond to stars and galaxies classified without any ambiguity. For probabilities far from 0 or 1, it presents some noisy measurements, coming mainly from faint galaxies. Figure \ref{heat-map} shows the same information but as a function of magnitude. \\

\begin{figure}
     \begin{center}
          \includegraphics[width=0.5\textwidth]{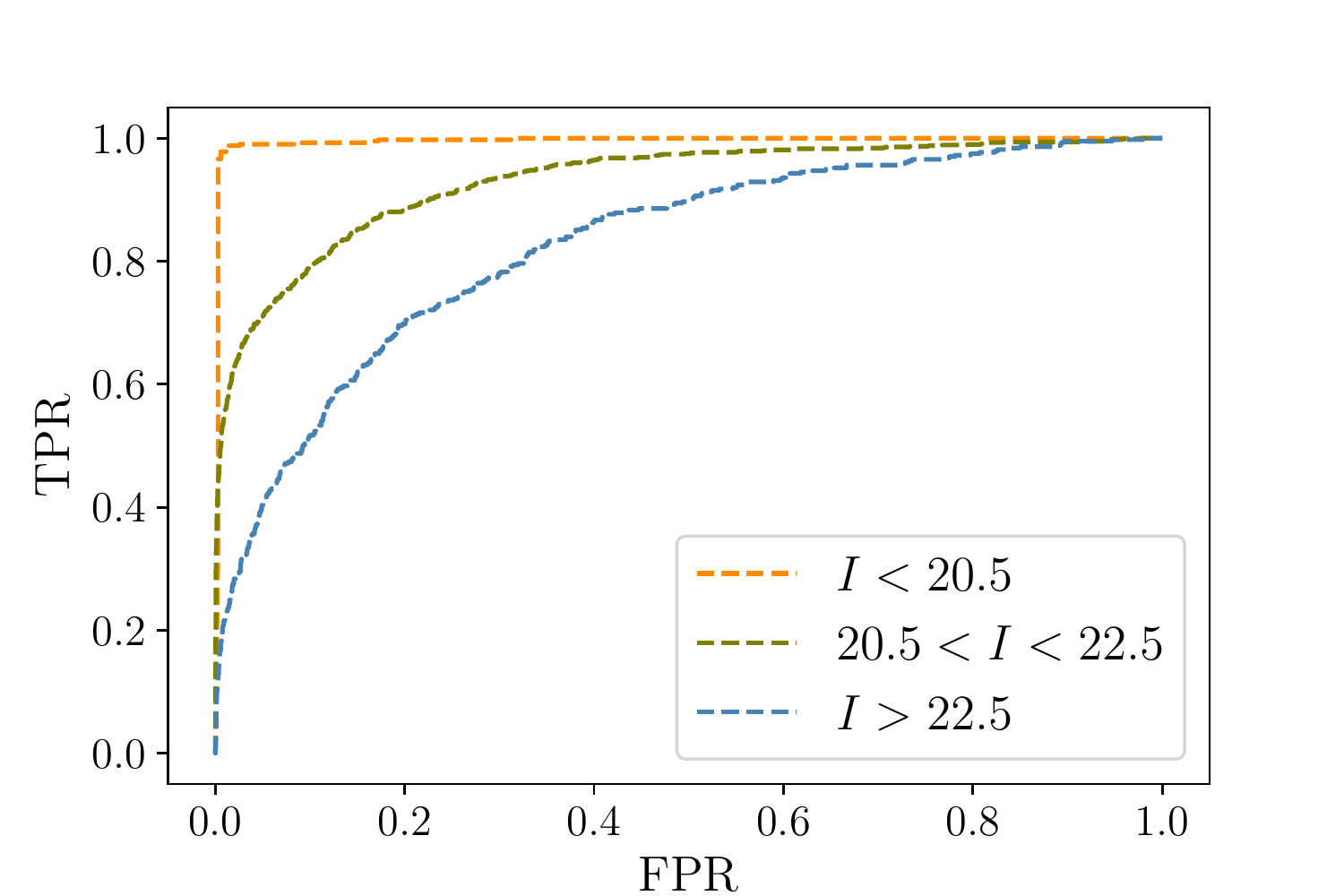}
      \end{center}
    \caption{ROC curve for star-galaxy classification on PAUS data given different cuts on the magnitude. The results plotted are those obtained on the validation sample.}%
   \label{PAU_magCuts}   
\end{figure}

Figure \ref{PAU_magCuts} shows the performance of the algorithm for training sets in three different magnitudes ranges: for $I<20.5$, $20.5<I<22.5$ and for $I>22.5$. \LC{As expected, it shows a degradation as the sample becomes fainter: giving ROC areas of 0.991, 0.930 and 0.822 for ranges $I<20.5$, $20.5<I<22.5$ and $I>22.5$, respectively}. Training sizes have been fixed to 3,500 objects for the three cases as the number of objects is limited by the smaller, brightest bin. Considering figure \ref{ROC_size}, the performances could still improve with a larger training set, specially for $20.5<I<22.5$ and $I>22.5$.\\

\begin{figure}
     \begin{center}
          \includegraphics[width=0.5\textwidth]{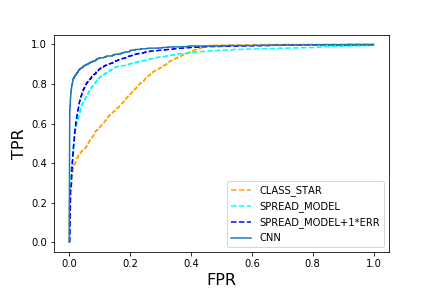}
      \end{center}
    \caption{ROC curves for star-galaxy classification in PAUS data using \cnn and \sxt classifiers. ERR corresponds to the \texttt{SPREADERR\_MODEL} quantity from \sxt. Both samples have been selected to have similar signal to noise distributions.}%
   \label{cnnvssex}   
\end{figure}

We can compare with morphological measurements on the same dataset, so a \sxt run was executed over the same field to obtain the \texttt{CLASS\_STAR} and \texttt{SPREAD\_MODEL} estimates of the shape of the object (see \citet{sgdes}). We used as an example the measurements in the 615 nm narrow band and compared with the \cnn results for a flux limited sample $I<22.5$ adjusting both samples to have the same signal to noise distributions. In Figure \ref{cnnvssex} we can see the advantages of using spectral information for classification, versus the standard morphological approach.  

\subsection{Classification on the ALHAMBRA catalog}

The application of the algorithm on the ALHAMBRA dataset should be useful to crosscheck the algorithm itself and also to test its power against an alternative classification scheme.  The ALHAMBRA survey also performed star-galaxy classification \citep{Molino} assigning a probability to every detection given its apparent geometry (the Full With at Half Maximum (FWHM) from \sxt, a synthetic F814W magnitude, and optical F489W - F814W and near infrared (NIR) J-Ks colors). The authors derived a probability distribution function (PDF) based on the typical distribution of stars and galaxies for each of the variables cited above. The final probabilities, the star-galaxy classifier, are included in the catalogs as the statistical variable \LC{\tt{Stellar Flag}}.\\

However, the ALHAMBRA images do not provide reliable morphological information for magnitudes $F814W > 22.5$, therefore the classification scheme is only applied up to this flux limit. For the rest of the catalog, they assigned a probability of 0.5. \\

It is of interest to see if by applying our algorithm based on low-resolution spectra on the ALHAMBRA catalog, \COMMENT{a multi-narrow band survey but with wider bands compared to PAUS,} we are able to match the purity provided (or even improve it) for objects brighter than $I = 22.5$. It is also of interest to see whether the algorithm is also able to classify faint objects for which ALHAMBRA did not provide any classification.\\


\LC{To get a classification for the ALHAMBRA catalog, we are entirely retraining the algorithm using ALHAMBRA's data. Instead of using fluxes, we are using isophotal AB magnitudes, checking the robustness of the classifier with a different range of inputs.}\\ 

The input features for the \cnn are a total of 23 parameters distributed as follows: the 20 mid-band optical magnitudes introduced as 19 colors, the 3 NIR broad bands $J,H,K$ magnitudes also included as 2 colors, the F814W magnitude and the Full Width at Half Maximum (FWHM).\\ 

The sample of objects for which ALHAMBRA also provided a classification (hence those with $F814W<22.5$) represents $20\%$ of the objects, with 5,500 galaxies and 1,500 stars, in the complete ALHAMBRA Gold catalog. Taking the COSMOS classification as the 'true' value for classification (admitting some QSO contamination), the ALHAMBRA classification obtains a ROC-AUC of 0.983.\\

\begin{figure}
	\centering
    \includegraphics[width=0.5\textwidth]{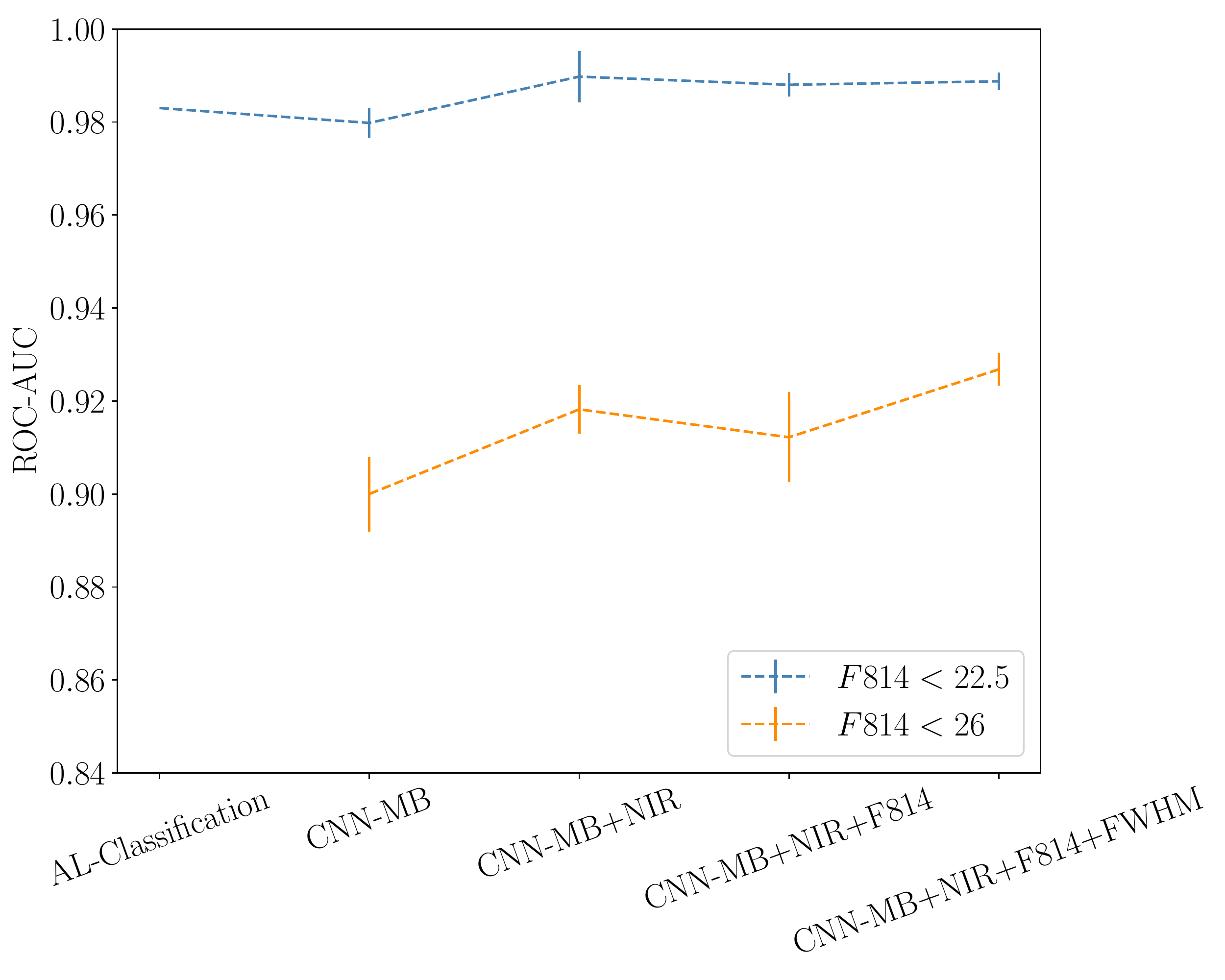}
    \caption{ROC AUC for the classification of the ALHAMBRA validation sample for different set of input features. In blue, classification for a object's sample with $F814W<22.5$ in orange, with $F814W<26$.}%
    \label{AL_class}
\end{figure}

Figure \ref{AL_class} shows the AUC-ROC of the classification we have performed on objects brighter than 22.5 (blue line) and objects brighter than 26 (orange line). There are different performances, each of them corresponding to the addition of new input features. Firstly, we have run the algorithm with only the optical band information. Then, we have added first the NIR information, then the F814W magnitude, \LC{where F814W refers to a synthetic broad band}, and finally the FWHM. Each line corresponds to the performance with a concrete \cnn feature set. This way, we can study how the algorithm scales as we add new features. The curve shows that by means of only the optical band information, the classification we get is similar to the original ALHAMBRA classifier \LC{\tt{Stellar\_Flag}}.\\ 

The addition of the NIR data makes the most difference and implies an important improvement in the classification performance (the power of the addition of infrared bands was already explored in \citep{banerji,kovacs,sgdes}). The best classification obtained is that with a ROC AUC of 0.99 corresponding to the performance with all the input features. The FWHM seems to be improving the classification only for fainter objects (orange line). However, it may be that the brighter ones already have a classification rate too high to be improved with an additional parameter. \LC{One can also notice that the additional F814W information is not improving the classification. The network's input already contains photometric spectral information with higher resolution than F814W and therefore, the broad band is not providing any extra information.} Table \ref{ALHAMBRA-AUC} contains the ROC-AUC values for the classification with the different input feature maps for both cases, brighter than 22.5 and 26.\\

\begin{table*}
\centering
\caption{ROC-AUC values for the classification of stars and galaxies in the ALHAMBRA dataset for different input feature maps. ALHAMBRA's \texttt{Stellar\_Flag} provides a ROC-AUC of 0.983 up to $F814W<22.5$.}
\label{ALHAMBRA-AUC}
\begin{tabular}{c|c|c|}
\cline{1-3}
\multicolumn{1}{c|}{\textbf{Information used}}                                            & \multicolumn{1}{l|}{\textbf{F814 \textless 22.5}} & \multicolumn{1}{l|}{\textbf{F814 \textless 26}} \\ \hline
\multicolumn{1}{|c|}{Optical Bands}                     & 0.980$\pm$0.003                                   & 0.901$\pm$0.001                                      \\ \hline
\multicolumn{1}{|c|}{Optical bands + NIR}               & 0.987$\pm$0.002                                   & 0.918$\pm$0.006                                      \\ \hline
\multicolumn{1}{|c|}{Optical bands + NIR + F814}        & 0.988$\pm$0.002                                   & 0.910$\pm$0.007                                      \\ \hline
\multicolumn{1}{|l|}{Optical bands + NIR + F814 + FWHM} & \multicolumn{1}{l|}{0.989$\pm$0.002}              & \multicolumn{1}{l|}{0.927$\pm$0.004}                 \\ \hline
\end{tabular}
\end{table*}

\begin{figure}
	\centering
    \includegraphics[width=0.5\textwidth]{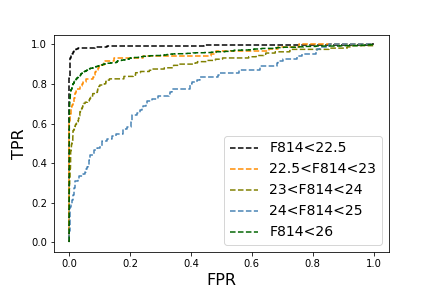}
    \caption{ROC curves obtained in the ALHAMBRA for different magnitude cuts.}%
    \label{AL_mag}
\end{figure}

As we did for PAUS (Figure \ref{PAU_magCuts}), Figure \ref{AL_mag} shows how the classification scales with the objects' magnitude.  For $F814W < 22.5$, we have already seen that the algorithm leads to a high ROC-AUC. 
Figure \ref{AL_mag} illustrates the performance of the algorithm in different binned magnitude ranges and, as expected, for magnitudes fainter than 22.5 the classification performance degrades. Nevertheless, considering all objects brighter than $F814W < 26$, we are able to obtain a \NS{$97.4\pm0.2\%$} purity for a completeness of \NS{$98.4\pm0.1\%$}.\\

As it was mentioned above, \NS{the ALHAMBRA catalogue also provides a classification (Stellar\_Flag)} for objects brighter than 22.5. For fainter objects, it is common practice to  consider every object to be a galaxy. This is also a good approach as there are relatively far fewer stars fainter than 22.5. 

\LC{Considering this approach, for $22.5<F814W<23$, the stellar contamination would represent an $8\%$ of the total dataset, with a $100\%$ completeness (by definition of the selection). However, with $cnn\_stellarity$ this contamination can be reduced to only $2.4\%$ for a completeness of $98.4\%$. Therefore, we significantly improve over this naive classification scheme with minimal loss. The same argument can be applied for fainter bins (keeping the same completeness levels) but with lesser gains in purity as the signal to noise decreases and the stellar sample becomes much smaller in relative terms. For $23<F814<24$, the contamination of stars is $4.8\%$ of the total dataset, whereas the algorithm achieves a $3.1\%$. Finally, within $24<F814<25$, the contamination of the sample would be $2.8\%$ for the naive classifier whereas we obtain $2.4\%$ using $cnn\_stellarity$. For fainter objects it is better to consider all sources as galaxies without performing any classification.}\\

\COMMENT{For $F814W<22.5$ the misclassified stars represent a $3\%$ of the total dataset, meaning that one would have a $3\%$ of contamination without nearly any misclassified galaxy. In the same magnitude range, we are able to obtain a purity of \NS{$97.8\%$}.\COMMENT{, hence it is worth to perform the classification.} If we move to $22.5<F814W<23$ and according to the COSMOS classification, considering all objects to be galaxies implies that the stellar contamination represents a $7.6\%$ of the total dataset. The \cnn is able to achieve a purity of $98\%$, therefore, we are able to improve this naive classification scheme for this magnitude range. \\
For $23<F814<24$, the contamination of stars is also  a $3.7\%$ of the total dataset, whereas the algorithm obtains a $97.5\%$ of purity. Finally, within $24<F814<25$, the contamination of the samples is of a $2.2\%$ whereas we get also a $98\%$ purity. Therefore, we are able to improve the ALHAMBRA classification up to $F814W<25$. For fainter objects on the other hand, it is better to consider all sources as galaxies. \\}

In order to further validate our algorithm, we have tested on a different field, ALHAMBRA-2, corresponding to DEEP2 observations \citep{Deep2}, training on ALHAMBRA-4 (COSMOS field). The reference catalog used here comes from matching to Hubble Space Catalog space imaging \citep{hsc} making a cut on extendedness of 1.2, which separates cleanly the point-like versus extended sources. \LC{The training and validation samples sizes are 20,000 and 5,000 objects respectively, all of them contained in ALHAMBRA-4}. For objects with F814W < 22.5, the \cnnc gives a ROC area of 0.943, while \textit{Stellar\_Flag} is 0.930. \LC{We also tested the remaining ALHAMBRA fields and obtained and updated classification, see appendix \ref{sec:Appendix}}.

\COMMENT{
\subsection{From broad bands to narrow bands: scaling with band width}

We have seen that a reliable star-galaxy classification based on the spectra is possible on both PAUS and ALHAMBRA. Although in the latter a morphological feature also plays a role in the classification, it is possible to get a high classification rate by means of the photometric spectra only.\\
In section \ref{sec:characterization}, we saw how the classification scales with the width of the bands joining consecutive bands together such that we obtain the same spectra represented by broader or narrower filter bands. \\
We will check this band width dependence comparing surveys with different photometric systems. We will compare three different datasets: one with the PAUS narrow bands, another with ALHAMBRA intermediate width bands and lastly a dataset from the Subaru telescope in $ugriz$ broad band filters, all of them over the COSMOS field. \\
To perform a fair comparison, we need to cover the same wavelengths for the three of them, in this case the PAUS wavelength range, from 450 nm to 850 nm. \\
Such range is covered with the 14 ALHAMBRA bands, from band F458W to band F861W, and with the Subaru $griz$ broad bands. We are going to consider objects with $S/N > 5$ in all cases. \\
Figure \ref{band_COMP} shows the comparison in the ROC space.  Again, one can see that the improvement is much larger from 4 to 14 bands than from 14 to 40. However, one could expect to have a better classification with 14 NB than with 14 MB \LC{... possible discussion on the StN}

\begin{figure}
	\centering
    \includegraphics[width=0.5\textwidth]{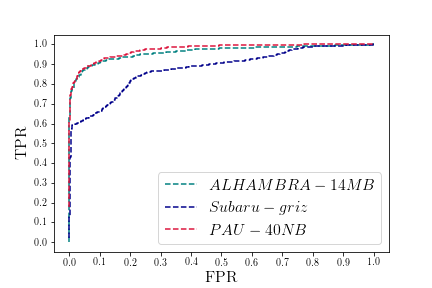}
    \caption{ROC curve for the classification of the PAUs data, the ALHAMBRA and the Subaru $griz$ with spectra over the same wavelength range.The results plotted are those obtained on the validation samples. }%
    \label{band_COMP}
\end{figure}
}

\section{Conclusions}
\label{sec:conclusions}
Convolutional neural networks have proven to be a real breakthrough in many fields, \LC{such as image pattern recognition, image classification, action recognition or document analysis}. We have shown here that they can be used as a powerful object classification tool using the shape of low resolution spectra from photometric data. \\

With such an algorithm, we have been able to classify stars and galaxies from the PAU survey by means of solely the object fluxes, without resorting to morphology, which in absence of a deeper detection image, can degrade significantly in the fainter end. This is done with a purity and a completeness of $99\%$ and  $98\%$, as shown in Figure \ref{PAU_magCuts}, using the COSMOS field as our training and testing grounds. \\ 

These results demonstrate the power of both the PAUS photometric quality, as the \cnn is able to detect subtle nuances in the spatial arrangement, such as characteristic of stellar spectra or emission lines, and use them to differentiate both populations. \\

In addition, using the same framework we have expanded and improved the ALHAMBRA classification. This survey also performed a star galaxy classification for objects brighter than 22.5, using magnitudes and also morphologies by means of a purely statistical method. We have applied our algorithm to their data with the same inputs, leading to a purity and a completeness of $98\%-99\%$. Adding the unclassified fainter objects, which contain the bulk of the catalog up to magnitude 26, leads to a purity of $97\%$ for a completeness of $99\%$ (nearly no misclassified galaxies). Under the assumption that all sources fainter than 22.5 are galaxies, we are able to improve the classification from objects with F814W < 25. This classification for the ALHAMBRA Gold catalog will be made available upon publication of this work.\\

\LC{The application of \cnn on low-resolution spectra from this kind of surveys opens up the possibilities beyond star-galaxy}
\COMMENT{The application of \cnn in such a fashion, using it to discover features in low-resolution spectra from this kind of surveys, opens up the possibilities beyond star-galaxy} classification, such as for the identification of other families of objects (e.g. adding a representative sample of quasars or AGNs in multi-labeled classification) or photometric redshift determination. This expands on their current astronomical applications which up to now where mainly for image processing and extraction of information from them directly.

An interesting avenue to explore is the comparison with template fitting methods. \NS{With the appropriate band information (blue and infrared)} they might prevail over machine learning methods in circumstances where the training set is poor \citep{Fadely}. Templates could additionally be used to augment the training set and improve classification when a wider range of labels is required.

\section*{Acknowledgements}
ISN would like to thank Miguel C\'ardenas-Montes, Iv\'an M\'endez and Ben Hoyle for useful suggestions and discussions concerning the use of CNNs. Funding for PAUS has been provided by Durham University (via the ERC StG DEGAS-259586), ETH Zurich, Leiden University (via ERC StG ADULT-279396 and Netherlands Organisation for Scientific Research (NWO) Vici grant 639.043.512) and University College London. The PAUS participants from Spanish institutions are partially supported by MINECO under grants CSD2007-00060, AYA2015-71825, ESP2015-88861, FPA2015-68048, SEV-2016-0588, SEV-2016-0597, and MDM-2015-0509, some of which include ERDF funds from the European Union. IEEC and IFAE are partially funded by the CERCA program of the Generalitat de Catalunya. The PAU data center is hosted by the Port d'Informaci\'o Cient\'ifica (PIC), maintained through a collaboration of CIEMAT and IFAE, with additional support from Universitat Aut\`onoma de Barcelona and ERDF. CosmoHub has been developed by PIC and was partially funded by the "Plan Estatal de Investigaci\'on Cient\'ifica y T\'ecnica y de Innovaci\'on" program of the Spanish government.





\bibliographystyle{mnras}
\bibliography{sgpau}


%
\begin{appendices}
\section{Appendix: The ALHAMBRA catalog extension with CNN classification}
\label{sec:Appendix}
As part of this work, we provide an additional column for the ALHAMBRA Gold dataset for which we have computed the stellarity value developed in this paper.

As training, we used the \citet{Leauthaud} dataset overlapping with ALHAMBRA-4, and we have updated the classification to cover objects up to $F814W<26.5$. The fields covered are from ALHAMBRA-2 to ALHAMBRA-8, in correspondence to DEEP-2, SDSS, COSMOS, HDF-N, GROTH, ELAIS-N1 and SDSS, respectively. The catalog with this classification is available at \url{http://cosmohub.pic.es}. In Table \ref{tab:catalog_columns} we provide the value added catalog columns that are being provided (most inherited from the original Gold catalog, for reference).\\

\begin{table*}
\centering
\caption{Description of the fields shaping the ALHAMBRA catalog where we provide $cnn\_stellarity$.}
\label{catalog}
\begin{tabular}{|c|c|}
\hline
\textbf{FIELD}    & \textbf{Objects}                                                                         \\ \hline
ID                & ALHAMBRA's objects unique identifier.                                                    \\ \hline
RA                & Right Ascension in decimal degrees.                                                      \\ \hline
DEC               & Declination in decimal degrees.                                                          \\ \hline
$Stellar\_Flag$   & ALHAMBRA's Statistical STAR/GALAXY Discriminator (0:Pure-Galaxy,0.5:Unknown,1:Pure-Star) \\ \hline
F814W             & Isophotal magnitude {[}AB{]}                                                             \\ \hline
dF814W            & Isophotal magnitude uncertainty {[}AB{]}                                                 \\ \hline
$cnn\_stellarity$ & CNN star/galaxy discriminator probability: {[}0:Pure-Galaxy,1:Pure-Star{]}               \\ \hline
\end{tabular}
\label{tab:catalog_columns}
\end{table*}

We only provide a classification for those objects with all bands measured (20 optical, 3 NIR and F814W). For those without, the class is set to a 'sentinel' value of -1.\\

The catalog is constructed training and validating on ALHAMBRA-4, in which `truth' classification is obtained from \citep{Leauthaud}, for those objects with the 24 bands measured. The training set counts with 13659 objects, whereas the validation set has 2096. Table \ref{obj_ALHAMBRA} shows the number of objects classified per field.\\

\begin{table}
\centering
\caption{Number of objects for which we have provided a classification per ALHAMBRA field.}
\label{obj_ALHAMBRA}
\begin{tabular}{|c|c|}
\hline
\textbf{FIELD} & \textbf{Objects} \\ \hline
ALHAMBRA-2     & 25856            \\ \hline
ALHAMBRA-3     & 27158            \\ \hline
ALHAMBRA-4     & 14946            \\ \hline
ALHAMBRA-5     & 15276            \\ \hline
ALHAMBRA-6     & 27400            \\ \hline
ALHAMBRA-7     & 26475            \\ \hline
ALHAMBRA-8     & 27813            \\ \hline
\end{tabular}
\end{table}
\end{appendices}

\COMMENT{\chapter{\large{\textbf{ ALHAMBRA catalog}}}\\ \
1. ID: Id number [integer]\\ \
2. X: Object position along x (pixel) [pixel]\\ \
3  Y: Object position along y (pixel) [pixel] \\ \
4. RA: Right ascension of barycenter (J2000) (deg) [real]\\ \
5. DEC:	Declination of barycenter (J2000) (deg) [real]\\ \
6. AREA: Isophotal area (pixel**2) [Integer]\\ \
7. FLUX from band 365 to 954: Isophotal flux (count) [real]\\ \
8. EFLUX from 365 to 954: Isophotal flux uncertainty (count) [real]\\ \
9. MAG from 365 to 954:	Isophotal magnitude	(AB) [real]\\ \
10. EMAG 365 to 954: Isophotal magnitude uncertainty (AB) [real]\\ \
11. FLUX for bands J,H,K: Isophotal flux (count) [real]\\ \
12. EFLUX for bands J,H,K: Isophotal flux uncertainty (count) [real]\\ \
13. MAG for bands J,H,K: Isophotal magnitude (AB) [real]\\ \
14. EMAG for bands J,H,K: Isophotal magnitude uncertainty (AB) [real]\\ \
15. FLUX F814: Isophotal flux (count) [real]\\ \
16. EFLUX F814:	Isophotal flux uncertainty (count) [real]\\ \
17. MAG F814: Isophotal magnitude (AB) [real]\\ \
18. EMAG F814: Isophotal magnitude uncertainty (AB) [real]\\ \
19. FLAGS:  Extraction flags [real]\\ \
20. CLASS STAR: \sxt S/G separation [real]\\ \
21. MASK SELECTION: Mask Selection [Boolean]\\ \
22. COLOR CLASS$\_$STAR: Color S/G Separation [real]\\ \
23. Z$\_$B: BPZ Photometric Redshift [real]\\ \
24. Z$\_$B$\_$MIN: Lower limit (95p confidence) [real]\\ \
25. Z$\_$B$\_$MAX: Upper limit (95p confidence) [real]\\ \
26. T$\_$B: BPZ SED type [real]\\ \
27. ODDS: BPZ Odds [real]\\ \
28. STELLAR MASS(log10($M\_sun$)): BPZ Stellar Mass [real]\\ \
29. M$\_$ABS: Absolute Magnitude (AB) ($B\_JOHNSON$) [real]\\ \
30. Z$\_$ML: Maximum Likelihood most likely redshift [real]\\ \
31. T$\_$ML: Maximum Likelihood most likely spect. type [real]\\ \
32. CHI SQUARED: Best fit Chi**2 [real]\\ \
33. Ks$\_$ABS: Absolute Ks Magnitude (AB) [real]\\ \

\chapter{\large{\textbf{The PAUS catalog}}}\\ \
1. Production$\_$id: Production ID number [integer]\\ \
2. Band: PAU bands \\ \
3. Flux: Forced aperture fluxes from NB455 to NB835 (count) [real]\\ \
4. Flux$\_$error: Forced aperture flux uncertainty (count) [real] \\ \
5. chi2: Reduced $\chi^2$ fit of the individual band measurements [real]\\ \
6. ra: Right Ascension in decimal degrees (deg) [real]\\ \
7. dec: Declination in decimal degrees (deg) [real]\\ \
8. $i\_auto$: \sxt MAG\_AUTO magnitude in the i band [real] \\ \
}
\COMMENT{
\chapter{\large{\textbf{CNN structure: scheme}}}\\ \
The \texttt{Keras} Python package \citep{keras} is a high-level neural network API that provides a set of tools to construct several machine learning codes. we have used \texttt{Keras} to construct our \cnn, that sketched in figure \ref{conv_structure}.\\
The CNN starts with a convolution-activation-pooling pattern. This is repeated three times with different kernel, pooling shapes and output dimensions. \\
The first convolution is done with a larger kernel, composed of 10 weights. The first pooling is also the largest, with size = 4. The activation function is in the three cases a 'Leaky-ReLu' function.
The algorithm first go through the more general traits of the spectra and then it reduces its the dimensionality, keeping the most important features.
In the two following convolution-activation-pooling groups, the kernel is reduced, looking for local features and the pooling is set to 2 in both cases. The two following layers consist on two fully connected layers (dense layers). There, are the neurons in a given layer are connected to that following it. The last dense layer has only two neurons, which are the number of outputs we want: the probability of being a star and that of being a galaxy.

\begin{figure}
	\centering
    \includegraphics[width=0.5\textwidth]{conv_structure.png}
    \caption{\cnn structure scheme}%
    \label{conv_structure}
\end{figure}
}


\bsp	
\label{lastpage}
\end{document}

%% file: authors.tex
\author[L.~Cabayol et al.]{L.~Cabayol$^{1}$\thanks{E-mail: lcabayol@ifae.es},
I.~Sevilla-Noarbe$^{2}$\thanks{E-mail: ignacio.sevilla@ciemat.es}, E.~Fern\'andez$^{1}$, 
J.~Carretero$^{1}$\thanks{Also at Port d'Informaci\'{o} Cient\'{i}fica (PIC), Campus UAB, C. Albareda s/n, 08193 Bellaterra (Cerdanyola del Vall\`{e}s), Spain},
M.~Eriksen$^{1}\footnotemark[3]$, 
\newauthor
S.~Serrano$^{3}$, 
A.~Alarc\'on$^{3,4}$,
A.~Amara$^{5}$,
R.~Casas$^{3,4}$,
F.~J.~Castander$^{3,4}$,
\newauthor
J.~de~Vicente$^{2}$,
M.~Folger$^{3,4}$, 
J.~Garc\'ia-Bellido$^{6}$, 
E.~Gaztanaga$^{3,4}$,
H.~Hoekstra$^{7}$, 
\newauthor
R.~Miquel$^{1,8}$, 
C.~Padilla$^{1}$,
E.~S\'anchez$^{2}$,
L.~Stothert$^{9}$, 
P.~Tallada$^{2}\footnotemark[3]$,
L.~Tortorelli$^{5}$\\ \\
$^{1}$ Institut de F\'{\i}sica d'Altes Energies (IFAE), The Barcelona Institute of Science and Technology, 08193 Bellaterra (Barcelona), Spain \\
$^{2}$Centro de Investigaciones Energ\'eticas, Medioambientales y Tecnol\'ogicas (CIEMAT), Madrid, Spain\\
$^{3}$Institute of Space Sciences (ICE, CSIC),  Campus UAB, Carrer de Can Magrans, s/n,  08193 Barcelona, Spain\\
$^{4}$Institut d'Estudis Espacials de Catalunya (IEEC), 08193 Barcelona, Spain\\
$^{5}$Institute for Particle Physics and Astrophysics, ETH Z\"urich, Wolfgang-Pauli-Str. 27, 8093 Z\"urich, Switzerland\\
$^{6}$Instituto de Fisica Teorica UAM/CSIC, Universidad Autonoma de Madrid, 28049 Madrid, Spain\\
$^{7}$Leiden Observatory, Leiden University, Leiden, The Netherlands\\
$^{8}$Instituci\'o Catalana de Recerca i Estudis Avan\c{c}ats, E-08010 Barcelona, Spain\\
$^{9}$Institute for Computational Cosmology, Department of Physics, Durham University, South Road, Durham DH1 3LE, UK\\
}